\documentclass[twocolumn,3p,times,procedia]{elsarticle}
\usepackage{ecrc}
\usepackage{graphicx}
\usepackage{titlesec}
\usepackage{kpfonts}
\usepackage{flushend}
\usepackage{amsmath,amsfonts,mathtools,epsfig}
\DeclarePairedDelimiter\abs{\lvert}{\rvert}%
\DeclarePairedDelimiter\norm{\lVert}{\rVert}%
\volume{3}
\firstpage{1}
\runauth{}
\usepackage{amssymb}
\usepackage{amsmath,ams}
\usepackage[figuresright]{rotating}
\usepackage{bm}
\usepackage{caption}
\usepackage{fancyhdr}
\fancyheadoffset[RO,EL]{0pt}
\usepackage{geometry}
\begin{document}

\begin{frontmatter}

\dochead{}
\title{
\begin{flushleft}
{\bf \Huge Windowed Overlapped frequency-domain Block Filtering Approach for Direct Sequence Signal Acquisition}
\end{flushleft}
}

\author[]{\bf \Large \leftline {Ebrahim Karami$^*$$^a$, Harri Saarnisaari$^b$}}

\address{\bf  \leftline {$^a$Department of Engineering and Applied Sciences, Memorial University, Canada, }

\bf  \leftline {$^b$Centre for Wireless Communications, University of Oulu, Finland}
}

\cortext[]{Ebrahim karami is responsible for all correspondance (email:ekarami@mun.ca).}

\begin{abstract}
This paper applies a windowed frequency-domain overlapped block filtering approach for the acquisition of direct sequence signals. The windows, as a novel viewpoint, not only allow pulse shaping without a front end pulse shaping filter, but also improve the performance of the spectrum sensing unit which can efficiently be implemented into this frequency-domain receiver and may further be used for spectrum sensing in cognitive radios or narrowband interference cancellation in military radios. The proposed receiver is applicable for initial time synchronization of different signals containing  a preamble. These signals include single carrier, constant-envelope single-carrier, multi-carrier and even generalized-multi-carrier signals, which makes the proposed receiver structure a universal unit. Furthermore, the receiver can be used to perform filtering with long codes and compute the sliding correlation of an unknown periodic preamble. It can further be modified to handle large Doppler shifts. We will also demonstrate the computational complexity and analysis of the acquisition performance in Rayleigh and Rician fading channels.
\end{abstract}


\end{frontmatter}

\section{Introduction}
Initial synchronization, or acquisition of a direct sequence (DS) signal appears to be a quite common first step that a communication receiver has to perform after switching the power on, because many wireless standards either use a DS signaling or their preamble, used for synchronization purposes, is a DS signal. These standards include GSM, LTE, UMTS, GPS, GALILEO, WIMAX, Zigbee, and many others wireless standards \cite{ergen2009,halonen2004,farahani2011,prasad2005}. For example, LTE systems use two DS signals, i.e., a 62-length Zadoff-Chu sequence and an 31-length M-sequence, as primary and secondary synchronization signals \cite{karami2015}. On the other hand, the performance of channel estimation and equalization, and data detection, algorithms is significantly affected by the accuracy of the initial synchronization \cite{karami2007blind,karami2004maximum,bennis2007performance,karami2013performance,karami2004joint,karami2006decision,karami2007equalization,estarki2007joint}. One solution to improve the synchronization robustness is to use interference cancellation (IC) signal processing \cite{karami2006very,karami2003new,karami2007near,karami2008near}. Notch filters are well-known examples of these. Another application for these IC units is spectrum sensing in cognitive radios. A notch filter may be a separate stand-alone unit in the front of a conventional receiver, but they may also be integrated into a frequency-domain receiver, which reduces the complexity, because the required transformations may be shared. frequency-domain receivers require less complexity and hence, have found many applications. One particularly interesting type of filtering is matched filtering, which allows fast acquisition \cite{torrieri2015,SIMON02}. In traditional frequency-domain filtering, where the filter is in a one piece,  overlap-save (OLS) or overlap-add (OLA) methods have to be acquired to properly handle the convolution process \cite{ingle2016}. Moreover, the frequency-domain receivers may be of interest in multipurpose or universal receivers, because they can be naturally used not only to receive multi-carrier signals such as orthogonal frequency division multiplexing (OFDM) and its variants as MC-CDMA \cite{pintelon2012} and generalized-multi-carrier (GMC) \cite{bica2016} signals but also to receive single-carrier signals \cite{hassanieh2012}.
\par
Some systems employ long DS codes and consequently require long filters which are difficult to implement \cite{karami2002efficient,karami2012novel}. In such cases, the filtering has to be divided into blocks and the required filtering process has to be performed using a process known as a block or partitioned filtering \cite{rao2014}. This technique is well-known in audio signal processing \cite{gay2012, smith2013}. Even overlapping blocks may be used \cite{KUK05, smith2013}. Block filters may also be adapted to acquire larger Doppler shifts than sole filters, see \cite{BETZ04,saarnisaari2008frequency,mohammadkarimi2017number,mohammadkarimi2015novel} for time-domain approach. Block filtering is equal to DFT filter banks (multi-rate filters) and linear periodic time varying (LPTV) filtering \cite{rao2014} but also short time Fourier transform (STFT)-based-filtering \cite{le2013}. The STFT adds windows, not used in DFT filter banks or LPTV filters, to the overall picture. The windows may be used to perform the pulse shape filtering, i.e., to match the filter frequency response to that of the signal and to improve the performance of notch filters by reducing the spectral leakage. However, although essential for proper performance of notch filters, windowing is known to cause signal-to-noise ratio (SNR) losses which are up to 3 dB for good windows. This loss may be reduced almost down to zero dB using overlapping segments, which are also elementary for STFT-based-processing \cite{CAPOZZA00}. A STFT-based-correlator DS-receiver is presented in \cite{QUYANG01}. In addition to data demodulation investigated in \cite{QUYANG01}, it may be used for serial search acquisition, which is known to result a slower acquisition than the matched filtering acquisition investigated herein.

This paper presents a frequency-domain, windowed, overlapped block filtering approach for DS signal acquisition. In addition to introducing the filtering and the acquisition concept, its other possible applications in radio communications will be briefly discussed. These include i) addition of a particular notch filter method \cite{SAARNISAARI05} into the receiver chain, ii) processing of different signals like conventional DS, constant envelope DS,  OFDM (WIMAX), MC-CDMA and GMC, iii) adaption the receiver to handle large Doppler frequency uncertainties and iv) possible changes when receiver is turned to the demodulation phase after acquisition. Furthermore, the paper includes analysis of computational complexity of the receiver compared to the conventional (non-block) matched filter implementation in the time or frequency-domain as well as analysis of acquisition probabilities in additive white Gaussian noise (AWGN) and Rayleigh flat fading channels, of which the latter are novel results. The probabilities include conventional detection and false alarm probabilities, maximum-search-based-probabilities and maximum search followed by threshold-detection-based-probabilities offering a very comprehensive picture of receiver's performance. These probabilities may be used to set the detection threshold and to predict the receivers performance in practice.  As a summary of this it could be said that the paper introduces a flexible baseband architecture that may be used with most existing and future signals and which offers spectrum sensing or narrowband interference rejection capability with a low additional cost. Therefore, the proposed receiver structure is a candidate receiver architecture for future multi-waveform platforms.

The rest of the paper is organized as follows. Section \ref{blockfiltering} introduces the filtering concept whereas applications and modifications are discussed in section \ref{applications}. The acquisition process is analyzed in section \ref{analysis} and simulation results confirming the analysis are shown in section \ref{simulations}. Finally, conclusions will be drawn in section \ref{conclusions}.
\section{Block Filtering}\label{blockfiltering}
This section first discusses block-wise convolution to provide an insight how the block filtering works and then present its mathematical frequency-domain basis, the STFT-based time-varying filtering.

\subsection{An Example}\label{convolution}
A simple example is probably the best way to explain how the block filtering differs from the conventional one. Let $x_1,x_2,x_3,x_4$ be the signal block to be filtered. In the conventional filtering, the signal is continuously fed into the filter whose impulse response is $h_1,h_2,h_3,h_4$. As a consequence, the response sequence is $x_1h_1, x_1h_2+x_2h_1, x_1h_3+x_2h_2+x_3h_1,x_1h_4+x_2h_3+x_3h_2+x_4h_1\, (\text{desired phase in acquisition}),\, x_2h_4+x_3h_3+x_4h_2,x_3h_4+x_4h_3,x_4h_4$. The block-wise convolution should end up to the same result.

In the block processing, the signal and the filter are divided into blocks using equal divisions. In the example, the division of the signal could be (the filter is divided correspondingly)
\begin{equation*}
\label{eq:example}
 \begin{bmatrix} x_3 & x_1 \\ x_4 & x_2\end{bmatrix},
\end{equation*}
where the block size $M=2$ and totality is $2\times 2$ matrix. In the absence of noise, the signal stream includes zero blocks in both sides. In other words, the signal matrix stream is
\begin{equation*}
\begin{matrix}
0 & x_3 & x_1 & 0 \\
0 & x_4 & x_2 & 0 \\
\end{matrix}.
\end{equation*}
This is divided into $2\times 2$ matrices by discarding the oldest data and taking a new block in. The input matrices, the first on right, are therefore
\begin{equation*} \begin{matrix}
\begin{bmatrix} x_1 & 0 \\ x_2 & 0 \\ \end{bmatrix} &  \begin{bmatrix} x_3 & x_1 \\ x_4 & x_2 \\ \end{bmatrix} & \begin{bmatrix} 0 & x_3 \\ 0 & x_4 \\ \end{bmatrix}
\end{matrix}.
\end{equation*}
In the block processing, only one input matrix is processed at each time instant, called as a filtering cycle. Each block (column) of an input matrix is convolved with the corresponding block (column) of a filter and the results are added together. Then, the next input matrix in the next cycle is received and the operations are repeated. Therefore, $M$ responses are calculated in one time cycle. To obtain the whole response, the operation has to be repeated for all $L$ possible cycles. Since the length of block convolution is $2M-1$, the tails have to be added to the corresponding convolutions in the next cycle. This is clarified next. It is assumed that each block (column) of the signal (matrix) passes a filter block from down to top. The corresponding convolution results are added together from each filtering cycle. The cycles are separated by bars and tails are below the dot lines. This results
\begin{equation*}
\begin{matrix}
  x_1h_1 & | & x_1h_3 + x_3h_1 & | & x_3h_3 \\
  x_1h_2+x_2h_1 & | & x_1h_4+x_2h_3 & | & x_3h_4+x_4h_3 \\
   & | & + x_3h_2+x_4h_1 & | &\\
  \ldots & | & \ldots  & | & \ldots \\
  x_2h_2 & | & x_2h_4 + x_4h_2 & | & x_4h_4 \\
           \end{matrix}.
         \end{equation*}
The tails of the convolution have to be added to the head of the convolution in the next cycle. More precisely, let $c_k=[h_h\ t_k]$ denote the convolution result in cycle $k$, where $h_k$ is the head (first $M$ samples) and $t_k$ the tail. In the next cycle $c_{k+1}=[h_{k+1}+t_k\ t_{k+1}]$. Therefore, the response of the block convolution becomes equivalent to the conventional convolution. As a summary: the signal stream is block by block fed through the filter, the column-wise convolution between the signal and the filter is performed, the convolution results are added column-wise together and the tails have to be added to the head of the next cycle. Since the convolution in the time-domain might be equally well performed  in the frequency-domain, in each cycle the FFT of the signal (matrix) can be element-wise multiplied by the FFT of the filter (matrix) and then is inverse transformed to obtain the time-domain convolution. After that, the convolution results are added together and OLA processing is performed. However, only one FFT per incoming signal block has to be calculated, because these transformations flow matrix-wise through the filter.
\par
By using a similar example, one can easily see that in the overlapping segments case (like $x_1,x_2;\, x_2,x_3;\, x_3,x_4$), the response of the block-wise convolution is not equal to the one of the conventional convolution. Instead, the original signal and its overlapped version have to be processed separately and the results have to be added afterwards. The filter has to be overlapped correspondingly.

\subsection{STFT-Based Block-Filtering}
All this is put into the STFT framework as follows. Let $x(n),\ n=0,\ldots,N-1$ be a discrete signal. Its STFT is \cite{le2013}
\begin{equation}
\label{eq:analysis}
X_{lm}=\sum_{n=0}^{N-1} x(n)w(n-lR)e^{j2\pi mn/M},
\end{equation}
where the analysis window $w(n)$ has length $M$ with non-zero values being in the interval $n=0,\ldots,M-1$. It is obvious that the signal is divided into blocks of $M$ samples and the blocks may overlap depending on the parameter $R$; if $R=M$ there is no overlapping, but just consecutive blocks. As a result of the analysis process \eqref{eq:analysis}, the signal is presented by a $M\times LM/R$ array of coefficients $X_{lm}$. For the simplicity, assume that $N=LM$ and $M/R=1,2,4,\ldots$. The case $M=R$ is called the critical sampling case. The selected restrictions yield to a simple implementation through FFT, but are  still quite flexible. More general case is studied in \cite{XIQI06}, but without considering signal acquisition.

There are several alternatives to recover the signal \cite{le2013}. One particularly interesting form is
\begin{equation}
\label{eq:synthesis}
x(n)=\sum_{l=0}^{L-1}g(n-lR)\sum_{m=0}^{M-1} X_{lm}e^{j2\pi mn/M},
\end{equation}
where $g(n)$ is the synthesis window of length $M$. Assuming that $w(n)$ and $g(n)$ satisfy some restrictions \cite{le2013}, the signal $x(n)$ can be perfectly reconstructed (synthesized) from its STFT coefficients $X_{lm}$. In other words, the STFT columns are first inverse-Fourier-transformed (rightmost sum in \eqref{eq:synthesis}), then windowed and finally added together in OLA fashion. Note that since the (I)FFT is a linear operator the order of addition and (I)FFT can be changed. Thus, if the synthesis window is rectangular, the complexity may be reduced performing addition before the IFFT. This naturally is a sensible operation, only if partial filtering results are not required like in Doppler processing or in filtering of several symbols during a filtering cycle.
\par
\begin{figure*}
\centering
\includegraphics[width=16cm]{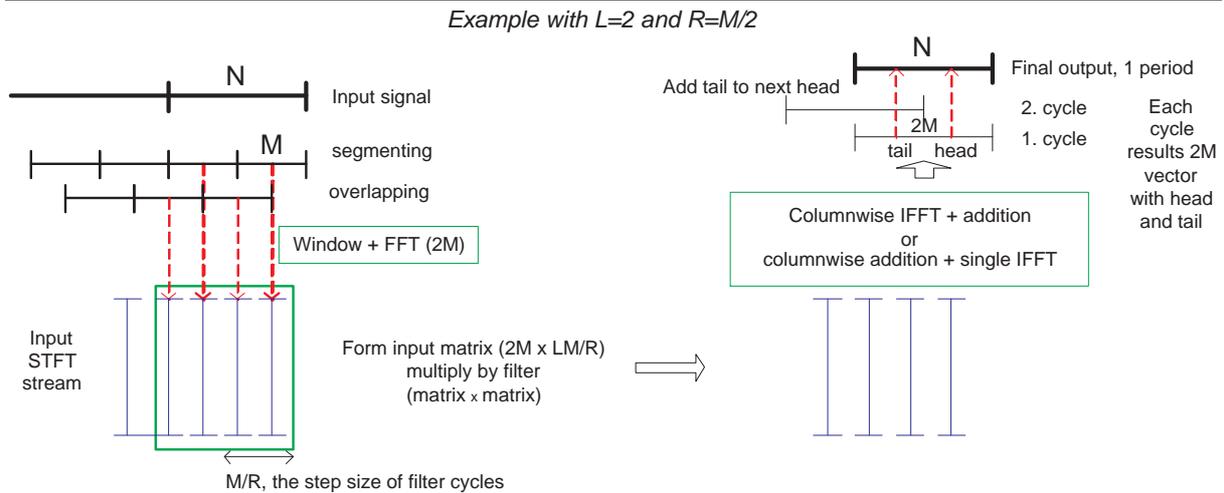}
\caption{An illustration of the windowed and overlapped block filtering approach.}\label{fi:STFTfilter}
\end{figure*}

Let $H_{lm}$ be the STFT of the filter.  It can be shown \cite{le2013} that the output of the filter is the inverse STFT of $X_{lm}H_{lm}$ (element-wise product). Thus, the filtering includes multiplication of the signal's STFT by that of the filter, and inverse transformation of the product. In the paper's case, the frequency response of the filter is zero outside an interval. Therefore, the output is computed as multiplying finite portion of signal's STFT with the filter's STFT. Furthermore, to handle the heads and tails properly, the FFT size has to be $2M$. The overlapping effect is taken into account by stepping the input signal STFT stream in steps of size $M/R$, the number of overlapping segments. The filtering process is illustrated in Fig. \ref{fi:STFTfilter}. Obviously, if $N=M=R$ the described block FFT filtering method reduces to the conventional FFT OLA filtering \cite{gay2012}.

\subsection{Complexity Comparison}
Herein, the complexity of generic time, conventional frequency-domain and block filtering are compared in terms of complex multiplications (CM). In the time-domain each output needs $N$ CM and there are $N$ outputs such that total complexity is $N^2$ CM. The conventional frequency-domain OLA processing needs FFT and IFFT of size $2N$ and multiplication by filters frequency response of size $2N$ yielding total complexity of $2N(\log_2N+1)$ CM.

The proposed block filtering (assuming rectangular windows) requires FFT of size $2M$, which has to be repeated $LM/R$ times, multiplying by filter of size $2M\times LM/R$ which has to be repeated $LM/R$ times. The conventional form still needs $LM/R$ IFFTs of size $2M$, whereas the simpler form has only $L$ IFFTs. Therefore, the total complexity of the conventional form is $\tfrac{2MN}{R}\big(\log_2 2M+\tfrac{LM}{R}\big)$ CM and that of the simpler form $\tfrac{2MN}{R}\big(\tfrac{1}{2}(1+\tfrac{R}{M})\log_2 2M+\tfrac{LM}{R}\big)$ CM. It can be observed that the complexity of the block filtering becomes equal than that of the conventional OLA filtering if $N=M=R$, as it should.

The complexity comparison results to a conclusion that the block filtering is more complex than the conventional one, but without overlapping the complexity increase is marginal. In addition, both frequency-domain versions are simpler than the generic time-domain implementation. However, possible windowing increases the complexity.

\section{Applications}\label{applications}
\subsection{Matched Filtering for Acquisition}
Symbol or chip synchronization is conventionally performed by correlation, but this results in a slow synchronization phase, see, e.g., \cite{torrieri2015}. A way to speed it up is to implement several correlators in parallel to simultaneously compute a number of test variables (search cells) \cite{BRAASCH07}. If the signal, to be synchronized, consist of $N$ symbols or chips, then the receiver usually has $qN$ search cells in time-domain, where $q$ is the oversampling factor. Additionally, there might be search cells in frequency as will be seen later but in this section only time uncertainty will be investigated. It is reminded that the receiver conventionally first includes a pulse shape filter whose response in fed into the correlator as one sample per symbol or chip basis. In the oversampling case, the response has to be split into $q$ steams and each stream is separately processed \cite{CHAPMAN01}. Alternatively, the pulse shape may be taken into account in the correlation \cite{MILLER06}. In this case, the receiver does not include a separate pulse shaping filter. This processing may be called as waveform-based-correlation, whereas the another processing may be called as training-symbol or chip-sequence-based-correlation. The STFT-based block-filtering may adopt both ways. In the former the analysis window, indeed, may be matched to form a suitable response. Another way to speed it up is to calculate the test variables through matched filtering either in time or frequency-domain \cite{SAARNISAARI04_PLANS2,BRAASCH07,MILLER06}.

In the serial search matched filter (MF) acquisition the outputs of the MF (computed in any possible way) are compared to a threshold in a serial fashion. In the maximum search a period of outputs is calculated and the maximum is found. This maximum is then compared the threshold. In both cases, it is claimed that the signal is present and symbol or chip synchronization has been acquired if the threshold is exceeded. This will be considered more detailed later in section \ref{analysis}.

\subsection{Different Modulations}
It is obvious that OFDM systems are a special case of the synthesized signal \eqref{eq:synthesis}, i.e., the window is rectangular and $L=1$. Conventional WIMAX synchronization is performed in time-domain. In WIMAX the preamble DS code is put into even subcarriers whereas odd ones are zero. This makes the time-domain signal periodic with two periods of size $N/2$, where $N$ is the number of subcarriers. The acquisition unit performs sliding correlation between two consecutive $N/2$ blocks \cite{SCHMIDL97}. There are also other variants and also a possibility that the matched filter and frequency-domain processing is used \cite{PUSKA07}. It should also be noticed that frequency-domain processing may be used to compute sliding correlation. However, sample by sample sliding results both in the time and frequency-domain to a high complexity. Therefore, the sliding step may be larger, e.g., $N/4$ or $N/8$. This is closely related to the overlap processing.

The generalized multi-carrier (GMC) transmission technique presented, e.g., in \cite{hassanieh2012,kliks2011,HUNZIKER03,XIQI06}, is a possible candidate for the future wireless communication systems. This is due to its better time-frequency localization properties which may reduce intersymbol and intercarrier interference, and remove need for the cyclic prefix needed in conventional OFDM systems \cite{bica2016}. The refereed papers consider different aspects of the GMC signal but not synchronization. In \cite{hassanieh2012} it was just mentioned that synchronization may be performed on subcarrier basis. The GMC signal may be explained as follows. The STFT coefficients $X_{lm}$ are the transmitted data symbols. In this case the signal \eqref{eq:synthesis} is called the GMC signal, or, if considered during interval $0,\ldots,N-1$, a GMC symbol corresponding the OFDM symbol definition. If a GMC system uses a known preamble symbol or symbols, its acquisition can be performed just as described here. By the authors knowledge, this is the first paper considering acquisition of GMC signals.

Linearly modulated single carrier signals can be obtained setting $M=1$. However, to keep the receiver universal one might to want to receive also these using the frequency-domain processing instead of conventional time-domain processing. This is possible since filtering, essential to all receivers, may be done either in the time or frequency-domain. This paper has readily shown how this is performed using frequency-domain block filtering. It is worth noting that also constant envelope DS signals may be received using a conventional matched filter \cite{BAIER84}, and thus the proposed frequency-domain block filter.

\subsection{Doppler Processing}
In the ideal Doppler processing, the input signal is transformed into different frequency offset corresponding to the possible Doppler values. In a simpler solution, the filter is divided (partitioned) into blocks and the outputs of these blocks are then Fourier transformed as shown in \cite{BETZ04} (and references therein). The reference uses time-domain processing, but as already shown, this partitioned matched filtering can be done also in the frequency-domain. If the Doppler is chancing (due to accelerated motion) between the blocks one may possible search over all possible (but sensible) Doppler tracks in the resulting time-frequency uncertainty grid. The acquisition probabilities concerning the Doppler processing are analyzed in \cite{BETZ04} and not repeated in this paper. Another way to increase Doppler resistance is to combine the partial responses either non-coherently or in a differentially coherent way \cite{MOON05}.

\subsection{Long Codes}
Some systems have a basic long code and direct implementation of a filter matched to it may be infeasible. Block filtering is a possible solution with shorter elements that are feasible to implement. Another case where block matched filters may be needed is when a long code is divided into subintervals and each subinterval contains a symbol. In this case responses of the partitioned matched filters are variables used for symbol demodulation (naturally sampled at symbol synchro position). This may be needed in a long code system where data rate is adjusted using code partitioning, but for some reasons short DS codes are not willed to be used.

A possible example where block filtering may be applied is the UMTS system where the uplink preamble consist of several scrambled repeats of a short code \cite{sesia2015}.

\subsection{Spectrum Sensing}
frequency-domain processing allows easy adaption of spectrum sensing algorithms since FFT is readily included into the processing chain. Spectrum sensing may be applied in cognitive radios to found available spectrum holes \cite{karami2011cluster}. Another application is interference cancellation (IC) needed especially in military systems. In these cases the process is known as notch filtering, but in both the cases the technique is basically the same. The window, inherent to the proposed receiver, is helpful since it reduces spectral leakage. However, a drawback of the windowing is the SNR loss, which may be 3 dB. Luckily overlapping, also inherent to the receiver, reduces this loss almost down to zero dB. 
\par
An important aspect, to notify when doing spectrum sensing or IC, is that if the desired underlaying signal is not flat, or white, in the frequency-domain also it may be detected (if SNR is high enough) or, what is worse, canceled. To avoid this unpleasant phenomena, the receiver should be designed using one sample per symbol/chip processing, i.e., the receiver should have a traditional pulse shaping filter at front and parallel processing of over-sampled streams. In this case we may loose an advantage of windows, but the complexity remains (almost) the same.

\subsection{Demodulation}
Once the acquisition is performed, the receiver turns its attention into tracking and data demodulation. In this turn the receiver may continue matched filtering if the signal has a DS component. The block filtering allows different code lengths; the short are needed at high data rates and the long are used in low data rates or when DS processing gain is needed for interference tolerance. In addition, the time varying nature \cite{le2013} of the filter allows de-spreading of scrambled signals. However, in this case the filter's or correlator's frequency response has to be updated frequently. Alternatively, the receiver uses correlation in the DS component case, pure FFT in the OFDM case or frequency-domain pulse shape filtering in the single carrier case. In the latter the filter may filter several symbols at one filtering cycle and the filter's frequency response is just the pulse shape. This pulse shaping goal may also be achieved  using a suitable analysis window.

\section{Acquisition Analysis}\label{analysis}
 One usually requires detectors insensitive to signal level variations called as constant false alarm rate (CFAR) detectors. These CFAR detectors may also be derived using generalized likelihood ratio detectors \cite{kay2013}. A CFAR detector is presented in \cite{SAARNISAARI04_ISSSTA,SAARNISAARI06-DS-FH}. Let $\vec{y}_k=a_k\vec{s}+\vec{n}_k$ denote the $k$th received signal including $N$ samples, where $a_k$ is a channel amplitude, $\vec{s}$ a preamble signal such that $\norm{\vec{s}}=1$ ($\norm{\ }$ denotes the Euclidean norm) and $\vec{n}_k$ a complex white Gaussian noise with variance $\sigma^2$. In addition, let $r(n)$ be an output of the MF (a test variable). If the signal is not present $a_k=0$. The detector is
\begin{equation}
\abs{r(n)}^2 > \gamma \norm{\vec{s}}^2\norm{\vec{y}_k}^2,
\end{equation}
where $\gamma$ is a parameter depending on the desired false alarm rate. The average signal power in the right hand side makes the detector a CFAR detector. It basically is an estimator of the thermal noise level, but it also makes the detector insensitive to interference. Note that if an IC unit is used, the average signal power has to be measured after the IC unit, i.e., after mitigation. This is so because mitigation may remove the interference that otherwise could deny detection. In other words, the mean signal power would be too high.

Another concern is that the effect of window on the threshold since it affects the signal power. This is more important, if input signal is windowed but the reference (filter) is not. Let $\vec{w}_a$ and $\vec{w}_r$ denote window vectors used for signal and filter analysis. Then, one has to use the power difference of windows as a normalizing factor. Furthermore, overlapping means that the computed response is replicated $M/R$ times. As a consequence,  the signal power should be modified as
\begin{equation}
\norm{\vec{s}}\equiv \frac{M}{R}\frac{\norm{\vec{w}_a}}{\norm{\vec{w}_r}}\norm{\vec{s}}.
\end{equation}
It can be shown \cite{SAARNISAARI04_ISSSTA,SAARNISAARI06-DS-FH} that the false alarm probability $P_{\text{FA}}$, i.e., the probability that the threshold is exceeded even though the signal is not present, can be approximated as
\begin{equation}
P_{\text{FA}}=e^{-\gamma N},
\end{equation}
from which $\gamma$ can  easily be obtained as $\gamma=\frac{1}{N}\ln(P_{\text{FA}})$, where . Another useful probability is the probability that the maximum exceeds the threshold. It can be shown to be \cite{SAARNISAARI04_ISSSTA,SAARNISAARI06-DS-FH}
\begin{equation}
P_{\text{FA,M}}=1-(1-P_{\text{FA}})^N.
\end{equation}

The probability of detection $P_{\text{D}}$, i.e., the probability that the test cell exceeds the threshold when the actual synchro position is investigated, can be approximated \cite{SAARNISAARI04_ISSSTA,SAARNISAARI06-DS-FH} as
\begin{equation}
\label{eq:PDawgn}
P_{\text{D}}=Q_0\big(\sqrt{2\mu},\sqrt{2\gamma(N+\mu})\big),
\end{equation}
where $\mu=\abs{a_k}^2/ \sigma^2$ is the signal-to-noise ratio (SNR) of the preamble signal and $Q_m(a,b)$ is the generalized Marcum Q-function \cite{ingle2016}.

Another useful probability is the probability $P_{\text{m}}$ that the maximum occurs at the actual synchro position. The approximation in \cite{SAARNISAARI04_ISSSTA,SAARNISAARI06-DS-FH} is not too accurate. Therefore, another attempt that will result a closer approximation is provided. Briefly explained, the analysis tool in  \cite{SAARNISAARI04_ISSSTA,SAARNISAARI06-DS-FH} considers the distribution of $r(n)$ as it is and assumes that $\norm{\vec{y}_k}^2$ converges to its average. This simplifies analysis since only one random variable has to be considered, but the method still has its roots on probability and statistics \cite{ROHATGI2015}. Now, at the synchro position $r(n)$ is a complex Gaussian variable with mean $a_k$ and variance $\sigma^2$. Thus,  $\abs{r(n)}^2$ has a non-central chi-square distribution. Assuming insignificant sidelobes on the autocorrelation function of the preamble signal, the non-synchro positions are zero mean Gaussian variables with variance $\sigma^2$. Now, the probability of interest is  $P_{\text{m}}=P(\abs{r_{\text{synchro}}(n)}^2\  > \abs{r_{\text{non-synchro}}(i)}^2, \ \forall i)$. A direct application of the analysis principle yields to result in \cite{SAARNISAARI04_ISSSTA,SAARNISAARI06-DS-FH}. However, this probability is equivalent the probability that the decision variable at the synchro position is larger than one of the largest non-synchro position. It is well-known that 98 \% of Gaussian variables are within 2.33 standard deviations from the mean. Thus, the novel approximation is
\begin{equation}\label{eq:pm}
P_{\text{m}}=Q_0\big(\sqrt{2\mu},\sqrt{2(2.33)^2}\big).
\end{equation}
For very long (large $N$) preamble signals the confidence probability may be higher, e.g., 99.5 \%, since it is natural that then, on average, the largest test variable at non-synchro positions may be larger than with short signals. See \cite{TURUNEN07} for another solution to this problem.

Still another probability of interest is the probability that the maximum exceeds the threshold independent of the fact is it the synchro position or not. This is \cite{SAARNISAARI06-DS-FH}
\begin{equation}
P_{\text{D,M}}=1-(1-P_{\text{FA}})^{N-1}(1-P_{\text{D}}).
\end{equation}
Finally, the probability that the maximum is at the synchro position and it exceeds the threshold is $P_{\text{M}}=P_{\text{m}}P_{\text{D,M}}$.

The above results are derived in additive white Gaussian noise (AWGN) case. In fading channels the situation is different. In Rayleigh fading channels, at the synchro position variable $r(n)$ follows a zero mean complex Gaussian distribution with variance $\sigma_s^2+\sigma^2$, where $\mu=E\{\abs{a_k}^2\}/\sigma^2=\sigma_s^2/\sigma^2$ is the average SNR, i.e.,
\begin{equation}
P(\abs{r(n)}^2)\equiv P(y)=1-e^{-y/(\sigma_s^2+\sigma^2)}.
\end{equation}
If the analysis tool in \cite{SAARNISAARI04_ISSSTA,SAARNISAARI06-DS-FH} is adopted, it follows that
\begin{equation}
P_{\text{D}}=e^{-\gamma N/(\mu+1)},
\end{equation}
whereas the paper's approach yields
\begin{equation}
P_{\text{m}}=\Big(e^{-(2.33)^2/(\mu+1)}\Big).
\end{equation}

In Rician fading channels, the decision variable of interest follows a complex Gaussian distribution with mean $a_k$ and variance $\sigma_s^2+\sigma^2$. Let the ratio of the power of the constant and random element be $\abs{a_k}^2/\sigma_s^2=\kappa$ and let $\mu=\abs{a_k}^2 / \sigma^2$ be the SNR of the constant element. Then, $\sigma_s^2+\sigma^2=\sigma^2(\mu/ \kappa+1)$. As a consequence,
\begin{equation}
P_{\text{D}}=Q_0\Big(\sqrt{\frac{2\mu}{\tfrac{\mu}{\kappa}+1}},\sqrt{\frac{2\gamma(N+\mu)}{\tfrac{\mu}{\kappa}+1}}\Big),
\end{equation}
which reduces to that \eqref{eq:PDawgn} in the AWGN channel (as it should) if the random element is weak since when $\sigma_s^2=0$, then $\kappa=\infty$. Correspondingly,
\begin{equation}
P_{\text{m}}=Q_0\Big(\sqrt{\frac{2\mu}{\tfrac{\mu}{\kappa}+1}},\sqrt{\frac{2(2.33)^2}{\tfrac{\mu}{\kappa}+1}}\Big).
\end{equation}

\subsection{More Exact Analysis}
This section provides more exact analysis of $P_{\text{m}}$ in AWGN, Rayleigh, and Rician channels.
\subsection{AWGN Channel}
Obviously, in an AGWN channel $P_{\text{m}}$ can be calculated as
\begin{equation}\label{eq:pm_awgn}
P_{\text{m}}^{AWGN}=Q_0\big(\sqrt{2\mu},\alpha_{N-1})\big),
\end{equation}
where $\alpha_{N}$ is defined as,
\begin{equation}
\alpha_{N}=\frac{E\{\max \abs{r(n)}^2, n=0,\ldots,N-1\} }{E\{\abs{r(0)}^2\} },
\end{equation}
where $E\{.\}$ is the expectation operator and $r(0)$ is the decision variable at the actual delay. The value of $\alpha_{N}$ closely follows a logarithmic function of $N$.

In the random channels the integral \eqref{eq:pm_awgn} has to be averaged over channel variations, i.e, integral
\begin{equation}
P_{\text{m}}=\int_{0}^{\infty} Q_0\big(\sqrt{2 \mu},\alpha_{N-1}\big) P(\mu)\, d\mu
\end{equation}
has to be solved, where $P(\mu)$ is the distribution of the SNR in a channel.

\subsection{Rayleigh Channel}
In a Rayleigh channel
\begin{equation}\label{eq:pm_rayl1}
P(\mu)=\frac{\mu}{\bar{\mu}} \exp(-\frac{\mu^{2}}{\bar{\mu}})
\end{equation}
where $\bar{\mu}$ is the average SNR. To solve the above integral, the generalized Marcum Q-function is replaced by its integral form. After some manipulations we will have
\begin{equation}\label{eq:pm_rayl2}
P_{\text{m}}^{Rayl}=(\frac{K \bar{{\mu}}}{K \bar{{\mu}}+1})^{1-K} \exp(-\frac{K \alpha_{N-1}}{K \bar{{\mu}}+1}),
\end{equation}
where $K$ is the number of PN sequences used for synchronization.

\subsection{Rice Channel}
In a Rician channel
\begin{equation}\label{eq:pm_rice1}
P(\mu)=\frac{\mu}{{\tilde{\mu}}} \exp(-\frac{\mu^{2}+\mu_{0}}{{\tilde{\mu}}}) I_{0}\big(\frac{\sqrt{\alpha_{N-1}}\mu}{\tilde{\mu}}\big),
\end{equation}
where $\tilde{\mu}$ is the average of the variable part of the SNR, $\mu_{0}$ is fixed part of the SNR such that $\bar{\mu}=\mu_{0}+\tilde{\mu}$, and $I_{0}(.)$ is the zero order modified Bessel function. To solve the needed integral, we have to replace the generalized Marcum Q-function with its equivalent integral form whereas the Bessel function is replaced by its Taylor series expansion. This series of integrals results
\begin{equation}
\begin{split}
P_{\text{m}}^{Rice}=&\sum_{n=0}^{\infty}\frac{2^{K-1}(K^2{\tilde{\mu}})^{n}}{(K {\tilde{\mu}}+1)}^{n+1}F(n+1,1,\frac{\mu_{0}}{2{\tilde{\mu}}(K {\tilde{\mu}}+1)})\\ &\cdot e_{n+K}(K\alpha_{N-1}),\label{eq:pm_rice2}
\end{split}
\end{equation}
where $F(.,.,.)$ is the hyper geometric function and $e_{n+K}(K\alpha_{N-1})$ is the incomplete exponential function defined as
\begin{equation}
e_{n+K}(K\alpha_{N-1})=\sum_{m=0}^{n+K-1}(K\alpha_{N-1})^{m}.
\end{equation}
Solution of \eqref{eq:pm_rice2} converges slowly. Convergence can be speed up by manipulating \eqref{eq:pm_rice2} into form
\begin{equation}
\begin{split}
P_{\text{m}}^{Rice}=&1-\sum_{n=0}^{\infty}\frac{2^{K-1}(K^2{\tilde{\mu}})^{n}}{(K {\tilde{\mu}}+1)}^{n+1}\\&F(n+1,1,\frac{\mu_{0}}{2{\tilde{\mu}}(K {\tilde{\mu}}+1)})\\& \cdot (\exp(K\alpha_{N-1}) -e_{n+K}(K\alpha_{N-1}).\label{eq:pm_rice3}
\end{split}
\end{equation}

\section{Numerical Results}\label{simulations}
 In this section the proposed windowed frequency-domain acquisition technique is simulated and then compared to the analytical results which provide bounds on the acquisition performance. As a reference, it is reminded that conventional non-windowed, non-overlapped approached achieve the theoretical bounds in the AWGN channel. Herein, it is trusted to a ``fact'' that if the analysis holds in Rayleigh fading channels it holds also in AWGN channels. Therefore, only Rayleigh fading channels are used in simulations. In all the simulations a 64 chips preamble sequence is used. It was a 63 chip Gold code extended by one. The signal is sampled one sample per chip. Simulation results are averaged over 1000 independent trials.  SNR is expressed per preamble sequence. The desired false alarm rate was quite high $10^{-2}$.

\begin{figure}
\centering
\includegraphics[width=1\linewidth]{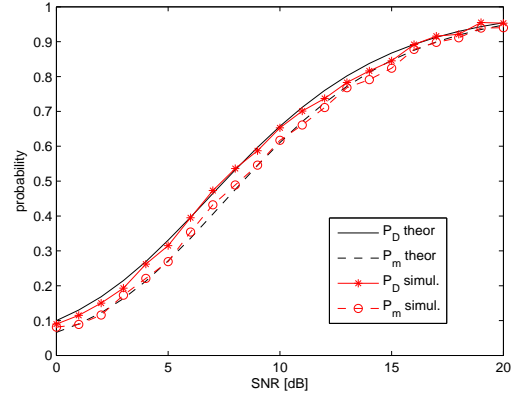}
\caption{Simulation and analysis results for a flat Rayleigh fading channel.}\label{Fig_Flat_rayl}
\end{figure}

Figure \ref{Fig_Flat_rayl} shows the results in a flat Rayleigh fading channel when $M=R=32$, i.e., overlapping is not used, and the window is rectangular. The results show that the simulated and theoretical results coincide, i.e., the approximative analysis is a proper one.

\begin{figure}
\centering
\includegraphics[width=1\linewidth]{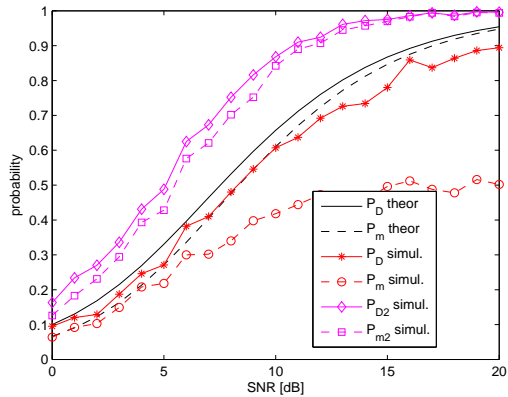}
\caption{Simulation results for a frequency selective Rayleigh fading channel. Theoretical values are for the flat fading channel.}\label{Fig_freqsel}
\end{figure}

Figure \ref{Fig_freqsel} shows interesting results concerning a frequency selective Rayleigh fading channel, which has two equal power multipath components with one chip separation. SNR is defined per path. In practice, the receiver does not know is the detected signal sample from the first or second path. Therefore, also probability that either the first or second path exceed the threshold ($P_{D2}$), and probability that either the first or second path provides the maximum ($P_{m2}$) are reported. It can be concluded from the results that diversity in the multipath channels is very beneficial for the synchronization. Of course, this benefit is lost if the second path is weak and situation becomes close to that in a single path channel. It can be seen that multipath propagation causes SNR losses to $P_D$. This is due to non-zero autocorrelation sidelobes, which are inversely proportional to the preamble length.  Another observation is that $P_m$ becomes close to half. This is easily understood since half of the time the second path is stronger than the first path if the paths have an equal power. It appears, although not shown in the figure for clarity reasons, that a good explanation of for $P_{i2}$, where $i$ is either $D$ or $m$, is
\begin{equation}
\label{eq:diversity}
P_{i2}=1-\prod_k \big(1-P_{i2}(\text{SNR}_k)\big),
\end{equation}
where the probabilities are expressed as a function of SNR and $\text{SNR}_k$ is the SNR of the $k$th path. This result follows from a though chain that probability that either the first or second (or $k$th) path exceeds the threshold is equal to probability that they all are below the threshold.

\begin{figure}
\centering
\includegraphics[width=1\linewidth]{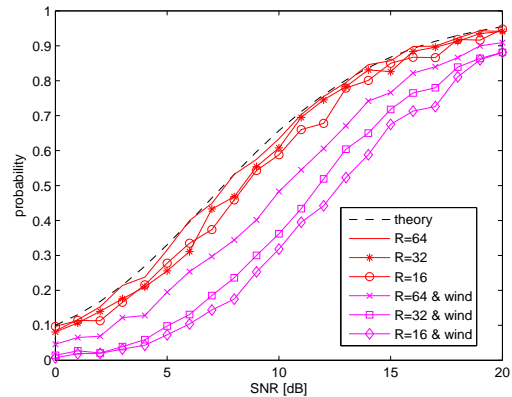}
\caption{Probability of detection $P_{D}$ simulation results for a flat Rayleigh fading channel when a window and overlapping are used.}\label{fi:wind_overlap}
\end{figure}

The last set of simulations concerns effects of windowing and overlapping. The used analysis window is the Kaiser window with the parameter 8, which has very low tail values. The window for the reference is rectangular. The overlapping is either non, 50 \% or 75 \%, i.e., $R=$ 64, 32 or 16 while $M=N=64$. The channel is a flat fading Rayleigh channel. The simulated false alarm rates with the original threshold setting are  0.01 (the desired value as it should), 0.16, 0.54, respectively, without the window and $4.7\times 10^{-6}$, $3.4\times 10^{-4}$, 0.05 with the window (640000 samples). This shows that threshold tuning is needed if a desired false alarm rate is needed with windows and overlapping. The trend seems to be that a non-rectangular window decreases the false alarm rate, whereas overlapping increases it. As a consequence, simulations with the original threshold setting would not be fair with respect the false alarm rate. Therefore, a proper threshold (multiplier of the original) was determined by simulations for overlapped and windowed cases. The results with equal false alarm rates are shown in Fig. \ref{fi:wind_overlap}. The results show that overlapping does not affect the performance significantly, but windowing does. Overlapping and windowing is even worse (by 2 dB) than just windowing the conventional non-blocked MF (M=N=R=64). The windowing loss with the conventional MF is  2--3 dB with this window.

\begin{figure}
\centering
\includegraphics[width=1\linewidth]{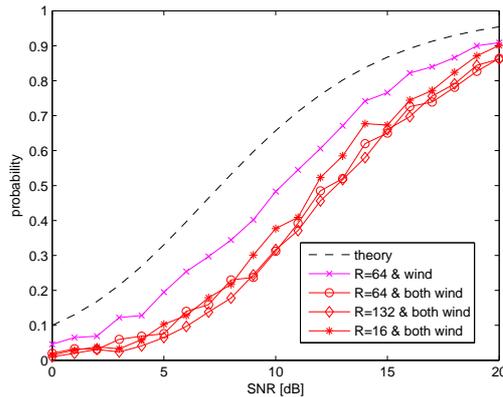}
\caption{Probability of detection $P_{D}$ simulation results for a flat Rayleigh fading channel when analysis and reference windows and overlapping are used.}\label{fi:wind_ovlap_rwin}
\end{figure}

The last result is contrary to the expectation that overlapping reduces windowing losses. Therefore, the last simulations use window also for the reference to see if that affects the situation. Fig. \ref{fi:wind_ovlap_rwin} demonstrates that adding of the reference window reduces the performance (decreases sensitivity), but now the overlapping does not further decrease it. The total loss compared to the theory is 5 dB. The results indicate that if the same sensitivity is required, then windowed cases have to have a higher false alarm rate. Maybe the mentioned expectation results from the fact that overlapping increases sensitivity if the threshold is kept constant. Therefore, paper's results might not be in contradiction to early ones.

\section{Conclusions}\label{conclusions}
The paper has provided insight into the windowed, overlapped, frequency-domain block filtering approach by explaining it and then showing (some of) its possible applications in radio communications. It was shown that this filtering approach may be used as a universal baseband receiver in communication systems, i.e., a single baseband architecture was shown to be able to receive all kind of signals. This is especially helpful in multipurpose platforms, which can (hereafter) be based on single architecture simplifying the design. Further investigations will be needed to see if this would reduce also other aspects in the receivers such as power consumption or silicon area.

In particular, the proposed approach was applied to signal acquisition with some novel analysis of acquisition probabilities in fading channels. This application and provided analysis and simulation results verify usefulness of the architecture for a wide range of the channel conditions. In addition, simulations showed that windowing reduces sensitivity if a desired false alarm rate is the receiver design goal. Therefore, one has to use windows with a care, e.g., in environments where they are really needed.

One future research topic with the proposed filter is that could a proper synthesis window be used to reduce the sensitivity losses the windows produce. Such a finding would improve usefulness of the filter. A way to find an answer might be the dual window. Another open question is the automatic detection threshold determination based on a given false alarm rate with overlapping blocks and windows.
\biboptions{numbers,sort&compress}
\bibliographystyle{elsarticle-num}
\bibliography{Paper}

\end{document}